\documentclass[a4paper,12pt]{article}
\pdfoutput=1

\usepackage{amsfonts,amsthm,amsmath,amssymb,graphicx,hyperref,cite,ytableau}
\usepackage[utf8]{inputenc}
\usepackage[all]{xy}
\usepackage[vcentermath]{youngtab}
\def\bea{\begin{eqnarray}} 
\def\eea{\end{eqnarray}}

\begin{document}
\makeatletter
\@addtoreset{equation}{section}
\makeatother
\renewcommand{\theequation}{\thesection.\arabic{equation}}
\vspace{1.8truecm}

{\LARGE{ \centerline{\bf Quantum Error Correction and Holographic}}}
{\LARGE{ \centerline{\bf  Information from Bilocal Holography}}}  

\vskip.5cm 

\thispagestyle{empty} 
\centerline{ {\large\bf 
Robert de Mello Koch$^{a,b,}$\footnote{{\tt robert.demellokoch@gmail.com}},
Eunice Gandote$^{b,}$\footnote{{\tt eunice@aims.edu.gh}}  }}
\centerline{{\large\bf
Nirina Hasina Tahiridimbisoa$^{b,c,}$\footnote{{\tt nirina@aims.ac.za}}   }}
\centerline{{\large\bf and
Hendrik J.R. Van Zyl${}^{b,}$\footnote{ {\tt hjrvanzyl@gmail.com}} }}

\vspace{.4cm}
\centerline{{\it ${}^{a}$ Guangdong Provincial Key Laboratory of Nuclear Science, Institute of Quantum Matter},}
\centerline{{ \it South China Normal University, Guangzhou 510006, China}}

\vspace{.4cm}
\centerline{{\it ${}^{b}$ National Institute for Theoretical Physics,}}
\centerline{{\it School of Physics and Mandelstam Institute for Theoretical Physics,}}
\centerline{{\it University of the Witwatersrand, Wits, 2050, } }
\centerline{{\it South Africa } }

\vspace{.4cm}
\centerline{{\it ${}^{c}$ Institute of High-Energy Physics of Madagascar (iHEPMAD)}}
\centerline{{\it University of Ankatso, Antananarivo 101, Madagascar,}}

\vspace{1truecm}

\thispagestyle{empty}

\centerline{\bf ABSTRACT}

\vskip.2cm 
Bilocal holography is a constructive approach to the higher spin theory holographically dual to $O(N)$ vector models.
In contrast to other approaches to bulk reconstruction, bilocal holography does not take input from the dual gravitational
theory. 
The resulting map is a complete bulk/boundary mapping in that it maps the complete set of $O(N)$ invariant degrees of 
freedom in the CFT, to the complete set of higher spin degrees of freedom.
After restricting to a suitable code subspace we demonstrate that bilocal holography naturally reproduces the quantum error
correcting properties of holography and it gives a robust bulk (entanglement wedge) reconstruction.
A gauge invariant entangled pair of CFT degrees of freedom are naturally smeared over a semicircle in the bulk spacetime,
which is highly suggestive of bit threads.
Finally, we argue that finite $N$ relations in the CFT, when interpreted in the dual AdS spacetime, can provide relations 
between degrees of freedom located near the boundary and degrees of freedom deep in the bulk.

\setcounter{page}{0}
\setcounter{tocdepth}{2}
\newpage
\setcounter{footnote}{0}
\linespread{1.1}
\parskip 4pt

{}~
{}~

\section{Introduction}

Holography, as embodied in the AdS/CFT correspondence\cite{Maldacena:1997re,Witten:1998qj,Gubser:1998bc}, is a 
profound new discovery about nature that is still poorly understood.
The correspondence describes a duality between a quantum gravitational theory in asymptotically AdS spacetime and 
a conformal field theory in one fewer spatial dimensions.
A question that has been surprisingly fruitful, goes under the name ``subregion duality''.
Subregion duality asks what region of the bulk AdS spacetime is dual to a subregion of the conformal field theory.
Questions of this type have lead to an appreciation that quantum error correction plays a role in 
holography\cite{Almheiri:2014lwa,Pastawski:2015qua,Harlow:2016vwg}, and also
to the notion of the holography of 
information\cite{Laddha:2020kvp,Banerjee:2016mhh,Penington:2019npb,Almheiri:2019psf,Almheiri:2019yqk,Penington:2019kki,Almheiri:2019qdq}, which adds to our understanding of
how information is coded into the bulk gravitational theory.
Our goal in this paper is to describe a simple setting in which these ideas can be probed and scrutinized.

We study the free $O(N)$ vector model in $2+1$ dimensions, described by the action
\bea
S=\int d^3 x \sum_{a=1}^N \left({1\over 2}\partial_\mu\phi^a \partial^\mu\phi^a\right)\label{cftaction}
\eea
This theory is holographically dual \cite{Klebanov:2002ja,Sezgin:2002rt}
 to higher spin gravity in AdS$_4$ \cite{Vasiliev:1990en,Vasiliev:2003ev,Didenko:2014dwa}.
As usual the single trace operators of the CFT give the single particle spectrum of the dual gravity.
The single trace operators consist of a single primary of spin 0 and dimension $\Delta=1$
\bea
O_{\Delta=1}(t,\vec{x})=\sum_{a=1}^N \phi^a(t,\vec{x})\phi^a(t,\vec{x})
\eea
and a tower of conserved higher spin currents, each of spin $s$ and dimension $\Delta=s+1$
\begin{eqnarray}
J_s(t,\vec{x},\alpha)
&=&J_{\mu_1\mu_2\cdots\mu_s}(t,\vec{x})\alpha^{\mu_1}\alpha^{\mu_2}\cdots \alpha^{\mu_s}\cr\cr
&=&\sum_{a=1}^N\sum_{k=0}^{s}
\frac{(-1)^k\, :(\alpha\cdot\partial)^{s-k}\phi^a(t,\vec{x})\;(\alpha\cdot\partial)^{k}\phi^a(t,\vec{x}) :}
{k!(s-k)!\Gamma(k+{1\over 2})\Gamma(s-k+{1\over 2})}
\cr
&&\label{scurrent}
\end{eqnarray}
where $\alpha^\mu$ is a polarization tensor that we can take to be null.
The three point functions of these spinning primaries reproduce the detailed prediction of the cubic
higher spin gravity \cite{Giombi:2009wh} at leading order in the gravitational coupling.

Bilocal holography \cite{Das:2003vw,Koch:2010cy,Jevicki:2011ss,Jevicki:2014mfa,Koch:2014mxa,Koch:2014aqa,deMelloKoch:2018ivk}, 
a constructive approach to holography, can be used to construct the holographic dual to this free CFT.
There are two basic ingredients that go into bilocal holography, both of which have a transparent physical motivation:
\begin{itemize}
\item[1.] Perform a change of variables in the path integral from the original field $\phi^a(t,\vec{x})$ to a 
gauge invariant bilocal field $\sigma(t,\vec{x}_1,\vec{x}_2)$. 
This change of field variable reorganizes the dynamics so that the loop expansion parameter for the bilocal
field theory is ${1\over N}$, matching expectations for the loop expansion of the dual holographic theory. 
\end{itemize}
The bilocal field given by 
\bea
\sigma(t,\vec{x}_1,\vec{x}_2)=\sum_{a=1}^N\phi^a(t,\vec{x}_1)\phi^a(t,\vec{x}_2) \label{bilocal}
\eea
is the complete set of equal time $O(N)$ invariant variables.
Notice that this change of field variables naturally leads to a higher dimensional theory: the original field $\phi^a(t,\vec{x})$
is defined in a $2+1$ dimensional spacetime, whereas the bilocal is defined in a $4+1$ dimensional spacetime.
This change of field variables has an important consequence: the original field $\phi^a(t,\vec{x})$ transforms in
a short irreducible representation $V_{{1\over 2},0}$ of the conformal group SO$(2,3)$ constructed on top of 
a primary field of dimension $\Delta ={1\over 2}$ and spin $s=0$.
In contrast to this, the bilocal fields transform in the tensor product $V_{{1\over 2},0}\otimes V_{{1\over 2},0}$.
The generators of conformal transformations, obtained by using the standard co-product, are representations of this tensor
product. The tensor product reduces as follows\cite{Flato:1978qz}
\begin{equation}
V_{{1\over 2},0}\otimes V_{{1\over 2},0} = V_{1,0}\,\oplus\,\,\bigoplus_{s=2,4,6,\cdots} V_{s+1,s}
\label{FlatoFronsdalintro}
\end{equation}
Each term in the direct sum on the right hand side corresponds to a field in the higher spin theory \cite{Vasiliev:1990en}.
Quite independent of this observation, it is natural to interpret each term in the direct sum as a distinct field.
To develop this point of view we should change basis for the Lie algebra of SO$(2,3)$.
This motivates the second step in bilocal holography
\begin{itemize}
\item[2.] Performs a change of variables from the CFT bilocal coordinates (which label points in a $1+4$ dimensional
spacetime) to coordinates for an AdS$_4\times$S$^1$ (which is also a $1+4$ dimensional spacetime). 
This change of coordinates takes us from a basis of generators (denoted collectively as $L_{\otimes}^A$) for SO$(2,3)$ 
naturally describing the LHS  of (\ref{FlatoFronsdalintro}) to a basis of generators (denoted collectively as $L_{\oplus}^A$
)naturally describing the RHS. 
\end{itemize}
This second point is highly non-trivial and is discussed in detail in the original works \cite{Koch:2010cy,Koch:2014aqa}, 
as well as in Appendix \ref{BMap}. 
The compact space S$^1$ which has made an appearance, turns out to be a nice way to organize the complete set of fields 
in higher spin gravity, very much in harmony with the ideas in \cite{Das:2003vw}.
We do not know a proof that the change of basis required by (\ref{FlatoFronsdalintro}) can always be accomplished with a
change of coordinates, but for the case of lightcone quantization \cite{Koch:2010cy} this is straight forward to demonstrate
as reviewed in Appendix \ref{BMap}.
In a light front quantization equal $x^+$ slices are used to define the Hilbert space.
The light cone coordinates are given by $x^\pm=x^1\pm x^0=x^1\pm t$. 
The coordinate $x^2$, which is transverse to the light cone will simply be denoted as $x$ in what follows.
The relevant bilocal is given by
\bea
\sigma (x^+,x_1^-,x_1,x_2^-,x_2) =\sum_{a=1}^N\phi^a(x^+,x_1^-,x_1)\phi^a(x^+,x_2^-,x_2)
\label{xplusbilocal}
\eea
These two steps completely determine the bilocal field theory.
For another approach, related to bilocal holography, see \cite{Aharony:2020omh}.

This construction of bilocal field theory is entirely motivated from the CFT and has not taken as input, any information
from the dual higher spin gravity.
It is interesting to ask if the bilocal field theory reproduces any features of the dual gravitational description.
In Section \ref{bulkreconstruction}, following \cite{Koch:2010cy,Koch:2014aqa}, we verify that the CFT equations of motion 
for the elementary field $\phi^a$ imply equations for the bilocal that reproduce the bulk equations of motion for the complete 
set of fields of higher spin gravity, and that the complete set of single trace CFT primaries are recovered from the boundary 
value of the bilocal field. 
This proves that bilocal holography, without any input from the dual higher spin gravity, achieves the desired bulk 
reconstruction.
Using this bulk reconstruction we are able to tackle the central question of which subregion of the bulk spacetime (if any) 
is dual to a given subregion of the conformal field theory? 
This is the subject of Section \ref{subregion}.
The subregion of the boundary CFT is obtained by restricting the coordinate $x$ transverse to the lightcone, to an interval.
Using bilocal fields living on this interval, we can reconstruct the bulk fields within a region of the bulk bounded by an 
extremal surface.
On the face of it, this appears to mesh nicely with ideas that have appeared in the entanglement wedge 
reconstruction \cite{Jafferis:2015del,Dong:2016eik,Cotler:2017erl,Chen:2019gbt}.
Entanglement wedge reconstruction tells us what part of the bulk is encoded in a given subregion of the boundary: only
bulk operators lying in the entanglement wedge of the boundary subregion can be reconstructed.
However, using the values of the fields that have been reconstructed, it is not possible to distinguish bulk fields with 
different spins.
To distinguish different spin components, even the reconstruction of fields at a single bulk point requires the complete 
boundary region.
There is no redundancy in how bulk information is coded into the boundary.
At the level of the complete CFT Hilbert space and the complete bulk AdS Hilbert space, AdS/CFT is a one-to-one
mapping between the states of the two Hilbert spaces.

This is puzzling since recent insights \cite{Almheiri:2014lwa,Pastawski:2015qua,Harlow:2016vwg} appear to have 
demonstrated and then exploited the fact that there is a certain redundancy in the bulk reconstruction.
An important ingredient in the quantum error correction approach to holography is the code subspace.
The code subspace is a construction that allows the description of effective field theory around a given classical background. 
Any low energy observer does not have access to infinite energy and they can act with, at most, a finite number of fields 
in the theory.
These limitations can be built in as cutoffs, on both the momentum and the occupation number.
In this way the Hilbert space which a low energy observer accesses can be constructed by acting on the relevant background 
state with a set of effective fields.
This formulation of effective field theory is state dependent \cite{Papadodimas:2013jku} since it requires a reference state. 
The resulting description is closely related to the construction of the little Hilbert space of states around black hole states given 
in \cite{Papadodimas:2012aq,Papadodimas:2013jku}, or equivalently, it is the code subspace of \cite{Almheiri:2014lwa}. 
For an explicit construction of the code subspace starting from an LLM reference state, see \cite{Berenstein:2017rrx}.
In Section \ref{qecsec} we consider how restricting to a finite energy code subspace affects our bulk reconstruction.
It is only after restricting the complete bulk AdS Hilbert space to the code subspace that reconstruction of local
bulk operators using a subregion of the boundary is possible.
In this case only a finite number of modes are excited, bulk information is encoded redundantly on the boundary and the
quantum error correction picture emerges.
The illustrative examples considered in \cite{Almheiri:2014lwa} find a concrete realization in the bilocal holography description.
Further, we find a natural emergence of the bit thread description of entanglement entropy developed in
\cite{Freedman:2016zud,Headrick:2017ucz,Cui:2018dyq}.
We are also able to demonstrate entanglement wedge reconstruction using bilocal holography.

Another puzzle which catalyzed significant progress is the monogamy problem: modes just inside and modes
just outside the horizon of a black hole must be entangled for a smooth horizon geometry.
Further, the Page curve demonstrates that modes just outside the horizon are entangled with modes far from the hole.
This is already a paradox because the monogamy of entanglement forbids entanglement between the modes just outside 
the horizon and both modes just inside the horizon and modes far from the hole.
One plausible resolution to this paradox, called the holography of information, claims that in fact the modes inside the 
black hole are not distinct degrees of freedom from the modes far from the black hole.
What is the mechanism for such an identification?
This holographic nature of the gravitational degrees of freedom only plays a role at late times when the state is a complicated 
state with many degrees of freedom at many different locations in spacetime (the emitted Hawking radiation) excited.
We argue that bilocal holography offers a natural mechanism for the holography of information: very complicated states
in the gravity have more degrees of freedom excited than there are independent degrees of freedom in the CFT.
This implies that there are relations between the gravitational degrees of freedom.
These are the analog of trace relations for matrix theories, except that here the relations are between degrees of freedom
at different locations in the bulk AdS spacetime.
We explain in particular, in Section \ref{HI}, that these relations naturally relate degrees of freedom localized at the 
boundary to degrees of freedom deep in the bulk AdS.
For ideas along these lines see also \cite{Jevicki:1996wu,Jevicki:1998rr,Aharony:2020omh}.
We present our conclusions and discuss our results in Section \ref{conclusions}.

Finally, note that bilocal holography is a special case of collective field theory \cite{Jevicki:1979mb,Jevicki:1980zg}.
The goal of collective field theory was always to change to invariant variables in order to obtain a loop expansion in 
${1\over N}$. 
This is a powerful insight which goes well beyond vector models.
In particular, we should expect an analogous construction for ${\cal N}=4$ super Yang-Mills theory.
The key obstacle is that there is not yet a manageable description for the invariant variables for matrix theories.
We will comment more on this possibility in Section \ref{conclusions}.

\section{Bilocal Bulk Reconstruction}\label{bulkreconstruction}

In this section we review bilocal holography as developed in 
\cite{Das:2003vw,Koch:2010cy,Jevicki:2011ss,Jevicki:2014mfa,Koch:2014mxa,Koch:2014aqa,deMelloKoch:2018ivk}.
The main novelty in our discussion is a mixed position space/momentum space description, which
will be useful when we discuss subregion duality in later sections.
Apart from this new representation, this section is a review of material developed in  
\cite{Das:2003vw,Koch:2010cy,Jevicki:2011ss,Jevicki:2014mfa,Koch:2014mxa,Koch:2014aqa,deMelloKoch:2018ivk}.
The key goal of this section is to explain how bilocal holography solves the bulk reconstruction problem.

A general formalism for changing to invariant variables in field theory, known as collective field theory, has
been developed in \cite{Jevicki:1979mb,Jevicki:1980zg}.
We use these general results in what follows. 
The change to bilocal variables is accomplished using the chain rule
\bea
\pi^a(t,\vec{x})&=&-i{\delta\over\delta \phi^a(t,\vec{x})}\cr
&=&-i\int d^2 y\int d^2 z \,\, {\delta\sigma(t,\vec{z},\vec{y})\over\delta \phi^a(t,\vec{x})}
\,\, {\delta\over\delta \sigma(t,\vec{z},\vec{y})}\cr
&=&\int d^2 y\int d^2 z\,\, {\delta\sigma(t,\vec{z},\vec{y})\over\delta \phi^a(t,\vec{x})}\, \Pi(t,\vec{z},\vec{y})
\eea
After changing to the bilocal variables (\ref{bilocal}) the CFT Hamiltonian
\bea
H=\int d^2 x \sum_{a=1}^N \left(
-{1\over 2}{\delta\over\delta\phi^a (\vec{x})}{\delta\over\delta\phi^a (\vec{x})}
+{1\over 2}\vec{\nabla}\phi^a\cdot\vec{\nabla}\phi^a\right)\label{cfthamiltonian}
\eea
becomes
\bea
H&=&2{\rm Tr} (\Pi\sigma\Pi)+{1\over 2}\int d^2 x \left(-\vec{\nabla}_x\cdot\vec{\nabla}_x\sigma (x,y)\right)\Big|_{x=y}\cr
&&+{N^2\over 8}{\rm Tr}(\sigma^{-1})-{N\over 2}{\rm Tr}(\sigma^{-1})\int dx\delta(0)
+{1\over 2}{\rm Tr}(\sigma^{-1})\left(\int dx\delta(0)\right)^2\label{collham}
\eea
where we are using a natural bilocal notation.
In this notation, for example,
\bea
{\rm Tr} (\Pi\sigma\Pi)=\int d^2x\,\,\int d^2 y\,\,\int d^2 z\,\, 
\Pi(t,\vec{x},\vec{y})\sigma(t,\vec{y},\vec{z})\Pi(t,\vec{z},\vec{x})
\eea
The second line in (\ref{collham}) arises from a non-trivial Jacobian resulting from the change of variables.
The Jacobian, which is a non-linear contribution, generates an infinite sequence of interaction vertices even for the free 
vector model.
The last two terms in (\ref{collham}) are counter terms that cancel divergences that appear when the ${1\over N}$
expansion is carried out.
The second and third terms of (\ref{collham}) define an effective potential
\bea
V={1\over 2}\int d^2 x \left(-\vec{\nabla}_x\cdot\vec{\nabla}_x\sigma (x,y)\right)\Big|_{x=y}
+{N^2\over 8}{\rm Tr}(\sigma^{-1})\label{effpot}
\eea
This potential plays an important role when the ${1\over N}$ expansion is developed.
We start by determining the ``classical'' large $N$ configuration $\sigma_0$ by solving
\bea
{\delta V\over\delta\sigma(t,\vec{x},\vec{y})}\Big|_{\sigma=\sigma_0} =0
\eea
We then expand about the leading solution as
\bea
\sigma(t,\vec{x},\vec{y})&=&\sigma_0(t,\vec{x},\vec{y})+{1\over\sqrt{N}}\eta(t,\vec{x},\vec{y})\cr
\Pi(t,\vec{x},\vec{y})&=&\sqrt{N}\pi (t,\vec{x},\vec{y})
\eea
It is the fluctuations $\eta(t,\vec{x},\vec{y})$ and $\pi(t,\vec{x},\vec{y})$ which are identified with bulk AdS fields.
They are a set of conjugate variables and obey the equal time commutation relation
\bea
[\eta(t,\vec{x},\vec{y}),\pi(t,\vec{x}',\vec{y}')]=i\delta(\vec{x}-\vec{x}')\delta(\vec{y}-\vec{y}')
\eea
This commutation relation is exact.
We know that the bilocal collective field theory will reproduce the CFT correlators to all orders in the $1/N$
expansion, since we have simply performed a change of variables.
This has been confirmed in detail in \cite{Jevicki:1983hb,deMelloKoch:1996mj}, by generating the loop expansion for the 
bilocal field theory and showing that it reproduces the $1/N$ expansion of the original vector model .
The non-trivial thing about the change of variables is that the loop expansion in the original variables is an
expansion in $\hbar$, whereas the loop expansion in the bilocal variables is an expansion in $1/N$.

The discussion above has been general.
Now we specialize to light front quantization.
The fluctuation around the bilocal field (\ref{xplusbilocal}) is denoted $\eta(x^+,x^-_1,x_1,x_2^-,x_2)$. 
Bulk AdS fields are identified with the fluctuation $\eta(x^+,x^-_1,x_1,x_2^-,x_2)$.
To spell out this mapping between bulk and boundary fields, we provide (i) a coordinate transformation between the 
coordinates of the bilocal field, and coordinates for the bulk AdS$_4\times$S$^1$ spacetime and (ii) an explicit mapping 
between the fluctuation $\eta(x^+,x^-_1,x_1,x_2^-,x_2)$ and the bulk AdS$_4\times$S$^1$ field 
$\Phi(X^+,X^-,X,Z,\theta)$. 
As mentioned in the introduction, this mapping is determined by the requirement that the basis for the Lie algebra is
transformed from the direct product basis appearing on the LHS of (\ref{FlatoFronsdalintro}) to the direct sum basis
appearing on the right hand side of (\ref{FlatoFronsdalintro}).
In this section we simply state the resulting map.
The reader wanting the details can find them in Appendix \ref{BMap}.

The coordinate transform between bilocal coordinates and those of AdS$_4\times$S$^1$ is conveniently developed in a mixed
position space/momentum space representation, obtained by Fourier transforming with respect to $x_1^-$ and $x_2^-$.
The bilocal field is now a function of $x^+$ as well as $p_1^+,x_1,p_2^+$ and $x_2$.
The bulk AdS spacetime has Poincare coordinates $X^0,X^1,X^2$ and $Z$.
Move to light cone coordinates $X^\pm =X^0\pm X^1$ and use the notation $X\equiv X^2$.
After Fourier transforming with respect to $X^-$, the AdS coordinates are $X^+,P^+,X,Z$.
There is also an angle $\theta$, which is a coordinate for S$^1$.
The coordinate transform relating the CFT coordinates $x^+,p_1^+,x_1,p_2^+,x_2$ to the AdS$_4\times$S$^1$
coordinates $X^+,P^+,X,Z,\theta$ identifies $x^+=X^+$ and
\bea
x_1&=& X+Z \tan \left(\frac{\theta }{2}\right)\qquad
x_2\,\,=\,\, X-Z \cot \left(\frac{\theta }{2}\right)\cr
p_1^+&=& P^+ \cos ^2\left(\frac{\theta }{2}\right)\qquad\quad
p_2^+\,\,=\,\, P^+ \sin ^2\left(\frac{\theta }{2}\right)\label{map1}
\eea
The inverse transformation is
\bea
X&=& \frac{p_1^+ x_1+p_2^+ x_2}{p_1^++p_2^+}\qquad
Z\,\,=\,\,\frac{\sqrt{p_1^+ p_2^+} (x_1-x_2)}{p_1^++p_2^+}\cr
P^+&=& p_1^++p_2^+\qquad\qquad
\theta\,\,=\,\,2 \tan ^{-1}\left(\sqrt{\frac{p_2^+}{p_1^+}}\right)\label{map2}
\eea
This map is non-trivial and it plays an important role in bilocal holography.
As we explain in Appendix \ref{BMap}, it is determined by CFT kinematics.
 
We will now explain the role of the coordinate $\theta$, which parametrizes an $S^1$.
This extra coordinate collects the complete set of four dimensional higher spin fields into a single five dimensional field.
The AdS$_4$ higher spin fields with spin $2s$, denoted as $\Phi_{2s}(X^+,X^-,X,Z)$, are obtained by developing 
$\Phi(X^+,X^-,X,Z,\theta)$ in an expansion as follows
\begin{equation}
\Phi(X^+,X^-,X,Z,\theta)=\sum_{s=0}^\infty\cos (2s\theta) \;\Phi_{2s}(X^+,X^-,X,Z)
\end{equation}
We are now ready so state the mapping between bulk and boundary fields.
The map between bilocal fields of the CFT and the higher spin bulk fields happens on a single time ($x^+$) slice
\begin{eqnarray}
&&\Phi\left(X^+,X^-,X,Z,\theta\right)=\int_{-\infty}^\infty dx_1^- \;\int_{-\infty}^\infty dx_2^-\; \; 2\pi i\sin\theta
\delta' (X^- -x_1^- \cos^2\frac{\theta}{2}-x_2^- \sin^2\frac{\theta}{2})\cr\cr
&&\qquad\qquad\qquad\qquad\qquad\times \eta\left(X^+,x_1^-,X+Z\tan {\theta\over2},
x_2^-,X-Z\cot{\theta\over 2}\right)
\end{eqnarray}
where the prime on the delta function denotes a derivative with respect to $X^-$.
We can rewrite this map in terms of the Fourier transformed bilocal
\begin{eqnarray}
&&\eta\left(X^+,x_1^-,X+Z\tan {\theta\over2},x_2^-,X-Z\cot{\theta\over 2}\right)
=\int {dp_1^+\over 2\pi}\int {dp_2^+\over 2\pi}e^{-ip^+_1 x_1^--ip^+_2 x_2^-}\cr
&&\qquad\qquad\qquad\qquad\qquad\times
\eta\left(X^+,p_1^+,X+Z\tan {\theta\over2},p_2^+,X-Z\cot{\theta\over 2}\right)
\end{eqnarray}
to obtain
\begin{eqnarray}
&&\Phi\left(X^+,X^-,X,Z,\theta\right)=\int_{-\infty}^\infty {dp_1^+\over 2\pi} \;\int_{-\infty}^\infty {dp_2^+\over 2\pi}
\; \; (2\pi)^2 (p_1^++p_2^+)\sin\theta e^{-i(p_1^++p_2^+)X^-}
\cr\cr
&&\qquad\qquad\qquad\qquad\qquad\times \delta (p_2^+\cos^2 {\theta\over 2}-p_1^+\sin^2 {\theta\over 2})\eta\left(X^+,p_1^+,X+Z\tan {\theta\over2},
p_2^+,X-Z\cot{\theta\over 2}\right)\cr\cr\label{niceJ}
&&
\end{eqnarray}
This is an off shell mapping between CFT fields and bulk AdS fields.
This interpretation is a simple consequence of the fact that the original integration variables for the path integral are the 
CFT fields $\phi^a$, while after the change of variables we integrate over the bilocals.
A simple rewriting gives
\bea
\Phi(X^+,p^+,X,Z,\theta) = 2\pi p^+\sin\theta\,\, 
\eta(X^+, p^+\cos^2{\theta\over 2},X+Z \tan{\theta\over 2},p^+\sin^2{\theta\over 2},X-Z\cot {\theta \over 2})\cr
\eea
The basic result proved in Appendix \ref{BMap}, says
\bea
L^A_{\oplus}\Phi(X^+,p^+,X,Z,\theta) = 2\pi p^+\sin\theta\,\, L^A_{\otimes} 
\eta(X^+, p^+\cos^2{\theta\over 2},X+Z \tan{\theta\over 2},p^+\sin^2{\theta\over 2},X-Z\cot {\theta \over 2})\cr
\eea
where the notation for the generators of the conformal group, $L^A_{\otimes}$ and $L^A_{\oplus}$ was defined in point
2 in the introduction.

Both the change of coordinates (\ref{map1}) and the operator mapping (\ref{niceJ}) are motivated entirely from the CFT.
Consequently, evidence that bilocal holography solves the bulk reconstruction problem is tremendously compelling.
The usual approach to reconstruction \cite{Banks:1998dd,Bena:1999jv,Hamilton:2006az}
constructs the bulk operators by perturbatively solving the bulk equations of motion, understood as operator equations in the
CFT, using the extrapolate dictionary \cite{Harlow:2011ke} to set boundary conditions.
We will now argue that, upon using the bilocal map, the CFT equations of motion imply the bulk equations of motion.
Further, the complete set of single trace primaries in the CFT are recovered in the $Z\to 0$ limit from the bulk AdS field
$\Phi(X^+,X^-,X,Z,\theta)$. 

The CFT equation of motion is the equation of motion for a free massless scalar
\begin{equation}
\left({\partial\over\partial x^+}{\partial\over\partial x^-}+{\partial^2\over\partial x^2}\right)\phi^a(x^+,x^-,x)=0
\end{equation}
This holds at the full quantum level, inside correlation functions, up to contact terms as usual.
After Fourier transforming to momentum space
\begin{equation}
\phi^a(x^+,p^+,x)=\int {dp^+\over 2\pi}\, e^{-ip^+x^-}\phi^a(x^+,x^-,x)
\end{equation}
the equation of motion is written as
\begin{equation}
i{\partial\over\partial x^+}\phi^a(x^+,p^+,x)={1\over p^+}{\partial^2\over\partial x^2}\phi^a(x^+,p^+,x)
\label{ffeom}
\end{equation}
Consequently, the leading large $N$ bilocal equation of motion, again up to contact terms, is
\begin{equation}
i{\partial\over\partial x^+}\sigma(x^+,p_1^+,x_1,p_2^+,x_2)=
\left({1\over p_1^+}{\partial^2\over\partial x_1^2}+{1\over p^+_2}{\partial^2\over\partial x_2^2}\right)
\sigma(x^+,p_1^+,x_1,p_2^+,x_2)
\end{equation}
Since this equation is linear, it must be obeyed by both $\sigma_0$ and $\eta$.
Consequently, we have
\begin{equation}
i{\partial\over\partial x^+}\eta(x^+,p_1^+,x_1,p_2^+,x_2)=
\left({1\over p_1^+}{\partial^2\over\partial x_1^2}+{1\over p^+_2}{\partial^2\over\partial x_2^2}\right)
\eta(x^+,p_1^+,x_1,p_2^+,x_2)\label{bilocaleom}
\end{equation}
Now, using (\ref{map1}) and (\ref{map2}) to change coordinates, it is a simple application of the chain rule to prove that
\begin{eqnarray}
&&i{\partial\over\partial x^+}\eta(x^+,p_1^+,X+Z\tan {\theta\over2},p_2^+,X-Z\cot {\theta\over2})\cr
&&={1\over p_1^++p_2^+}\left({\partial^2\over\partial X^2}+{\partial^2\over\partial Z^2}\right)
\eta(x^+,p_1^+,X+Z\tan {\theta\over2},p_2^+,X-Z\cot {\theta\over2})
\end{eqnarray}
Using the above CFT equation of motion, it is simple to see that
\begin{eqnarray}
&&{\partial\over\partial X^+}{\partial\over\partial X^-}
\Phi\left(X^+,X^-,X,Z,\theta\right)=-\int_{-\infty}^\infty dp_1^+ \;\int_{-\infty}^\infty dp_2^+
\; \sin\theta (p_1^++p_2^+)
e^{-i(p_1^++p_2^+)X^-}\cr\cr
&&\qquad\qquad\times\delta (p_2^+\cos^2 {\theta\over 2}-p_1^+\sin^2 {\theta\over 2})
\left({\partial^2\over\partial X^2}+{\partial^2\over\partial Z^2}\right)\eta\left(x^+,p_1^+,X+Z\tan {\theta\over2},
p_2^+,X-Z\cot{\theta\over 2}\right)\cr
&&\qquad\qquad
=-\left({\partial^2\over\partial X^2}+{\partial^2\over\partial Z^2}\right)\Phi\left(X^+,X^-,X,Z,\theta\right)
\end{eqnarray}
so that
\begin{eqnarray}
\left({\partial\over\partial X^+}{\partial\over\partial X^-}+
{\partial^2\over\partial X^2}+{\partial^2\over\partial Z^2}\right)\Phi\left(X^+,X^-,X,Z,\theta\right)=0
\end{eqnarray}
This is the equation of motion obtained for an arbitrary symmetric higher spin field in light cone gauge in 
AdS$_4$ \cite{Metsaev:1999ui}.

To complete the demonstration of bulk reconstruction, we must show that the complete set of single trace primaries can be
obtained from  the boundary behavior $Z\to 0$ of the bulk fields $\Phi_{2s}\left(X^+,X^-,X,Z\right)$.
A simple computation gives
\begin{eqnarray}
\Phi_{2s}\left(X^+,X^-,X,Z\right)\Big|_{Z=0}&=&
{2\over \pi}\int_0^\pi d\theta \cos (2s\theta)\Phi\left(X^+,X^-,X,0,\theta\right)\cr
&=&\int_0^\pi d\theta \cos (2s\theta)\int_{-\infty}^\infty {dp_1^+\over 2\pi} \;\int_{-\infty}^\infty {dp_2^+\over 2\pi}
\; \; 8\pi(p_1^++p_2^+)\sin\theta e^{-i(p_1^++p_2^+)X^-}
\cr\cr
&&\times \delta (p_2^+\cos^2 {\theta\over 2}-p_1^+\sin^2 {\theta\over 2})
\eta\left(X^+,p_1^+,X,p_2^+,X\right)
\end{eqnarray}
Using the delta function to perform the integral over $\theta$, we find
\begin{eqnarray}
\Phi_{2s}\left(X^+,X^-,X,0\right)&=&
\int_{-\infty}^\infty {dp_1^+\over 2\pi} \;\int_{-\infty}^\infty {dp_2^+\over 2\pi}
\cos \left(4s \arctan\sqrt{p_2^+\over p_1^+}\right)
\; \; 16\pi e^{-i(p_1^++p_2^+)X^-}
\cr\cr
&&\eta\left(X^+,p_1^+,X,p_2^+,X\right)\cr
&=&16\pi\;
\int_{-\infty}^\infty {dp_1^+\over 2\pi} \; e^{-ip_1^+X^-} \phi^a(X^+,p_1^+,X)\;\times\cr
&&\int_{-\infty}^\infty {dp_2^+\over 2\pi}\; e^{-ip_2^+X^-} \phi^a(X^+,p_2^+,X)\;
\cos \left(4s \arctan\sqrt{p_2^+\over p_1^+}\right)
\end{eqnarray}
Now, we note that
\begin{equation}
(p_1^++p_2^+)^s\, \cos \left(2s \arctan\sqrt{p_2^+\over p_1^+}\right)
=\Gamma \left(\frac{1}{2}\right) s! \Gamma \left(s+\frac{1}{2}\right)
\sum_{k=0}^s
\frac{(-1)^k (p_1^+)^{s-k}(p_2^+)^k }
{\Gamma \left(s-k+\frac{1}{2}\right) \left(\Gamma \left(k+\frac{1}{2}\right) k! (s-k)!\right)}
\end{equation}
which implies that
\begin{eqnarray}
{\partial^s\over \partial X^-{}^s}\Phi_{s}(X^+;X^-,X,0)
=16\pi\Gamma \left(\frac{1}{2}\right) s! \Gamma \left(s+\frac{1}{2}\right)\sum_{k=0}^{s}
\frac{(-1)^k \partial_-^{s-k}\phi^a(X^+,X^-,X)\partial_-^{k}\phi^a(X^+,X^-,X) }
{\Gamma \left(s-k+\frac{1}{2}\right) \Gamma \left(k+\frac{1}{2}\right) k! (s-k)!}\cr
\end{eqnarray}
We have recovered the conserved higher spin current of spin $s$, given in (\ref{scurrent}).
For $s=0$ we simply obtain the $\Delta=1$ scalar of spin zero.
This gives a primary state from each single trace conformal multiplet.
The complete set of states in the multiplet can be obtained by acting with generators of the conformal
algebra.
This demonstrates that the complete set of single trace primaries is obtained from  the boundary behavior $Z\to 0$ of the 
bulk fields $\Phi_{2s}\left(X^+,X^-,X,Z\right)$.

The equation of motion for the original CFT field (\ref{ffeom}), when applied to a bilocal field within a correlator, 
naturally gives rise to contact terms that encode corrections that are subleading at large $N$.
After changing variables to the bilocal field, the subleading terms are generated by an infinite sequence of vertices 
that arise upon expanding the Jacobian about the leading configuration.
These interactions should be reproduced by a fully gauge fixed version of Vasilliev's higher spin 
theory \cite{Vasiliev:1990en,Vasiliev:2003ev,Didenko:2014dwa}.
This seems hard to test explicitly, given the complexity of Vasiliev theory.
However we know that the vertices generated by the Jacobian are exactly the interactions needed to reproduce the 
correlators of the CFT, to all orders in ${1\over N}$, because correctly changing integration variables in the path integral 
does not change the value of correlation functions.
Assuming the vector model/higher spin duality, this is evidence that the ${1\over N}$ correction to the bilocal equations of 
motion will agree with the higher order corrections to the higher spin equations of motion, extending agreement beyond 
the leading order at large $N$. 
In this sense, the reconstruction of bilocal holography goes well beyond that of 
\cite{Banks:1998dd,Bena:1999jv,Hamilton:2006az} which is perturbative by construction.
This completes our demonstration that bilocal holography, without any input from the dual higher spin gravity, 
achieves the correct bulk reconstruction.

\section{Subregion duality}\label{subregion}

Which subregion of the CFT (if any) is dual to a given subregion of the bulk spacetime?
This deceptively simple question has motivated enormous progress: it lead to the discovery that there is some arbitrariness in 
the mapping between the CFT and the dual AdS theory, elegantly described as the statement that the bulk-to-boundary map 
in AdS/CFT defines a quantum error correcting code \cite{Almheiri:2014lwa,Pastawski:2015qua,Harlow:2016vwg}. 
This discovery was an important ingredient needed to answer the question posed by subregion 
duality \cite{Dong:2016eik,Cotler:2017erl,Faulkner:2017vdd,Chen:2019gbt}: the bulk region encoded
in an arbitrary boundary region ${\cal R}$ is the entanglement wedge of ${\cal R}$\cite{Hubeny:2007xt}. 
The entanglement wedge of boundary region ${\cal R}$ is the bulk region located in a single time slice, bounded by
${\cal R}$ itself and its Ryu-Takayanagi (RT) surface \cite{Ryu:2006bv,Ryu:2006ef}, which is the minimal area bulk 
surface anchored to the boundary of ${\cal R}$. 
Entanglement wedge reconstruction\cite{Dong:2016eik,Cotler:2017erl,Faulkner:2017vdd,Chen:2019gbt,Czech:2012bh,Headrick:2014cta,Wall:2012uf,Jafferis:2015del,Hayden:2018khn} provides a boundary operator acting only on the boundary subregion 
${\cal R}$ for each bulk operator acting on the entanglement wedge of ${\cal R}$. 
In this section using bilocal holography, we consider the subregion duality problem.
We focus on the subregion defined by restricting $x$ to an interval.
Bilocal fields restricted to this interval correspond to bulk fields within the entanglement wedge of the interval.  
These bulks fields are a linear combination of AdS$_4$ fields, with arbitrary even integer spin.
As we explain in detail, to distinguish the different spin components, even the reconstruction of fields at a single bulk 
point requires the complete boundary region.
Consequently, using the values of the fields that have been reconstructed, it is not possible to distinguish bulk fields with 
different spins.
At the level of the complete CFT Hilbert space and the complete bulk AdS Hilbert space, AdS/CFT is a one-to-one
mapping between the states of the two Hilbert spaces.
There is no redundancy in how bulk information is coded into the boundary.
In the next section we reconsider the subregion duality problem, after properly accounting for the
limitations faced by any low energy observer.
In this case the map provided by bilocal holography does indeed define a quantum error correcting code.

Our strategy entails considering localized CFT excitations and determining where they map to in the dual bulk spacetime.
Given the form of the bilocal map, it is most instructive to consider bilocal excitations at time $x^+$, with the first 
excitation localized at ($x_1,p_1^+$) and the second at ($x_2,p_2^+$).
The two excitations in the bilocal are described as wavepackets, tightly peaked at $x_1$ and $x_2$ along the spatial 
direction $x$ transverse to the light cone, but because they have a definite value for $p^+$, completely smeared along 
the  $x^-$ direction.
The map (\ref{map1}) - (\ref{map2}) mixes spacetime coordinates and momenta, so its not entirely obvious where
the CFT excitations are located in the bulk. 
Note the easily verified identity
\begin{equation}
\left(X-{x_1+x_2\over 2}\right)^2+Z^2 = \left({x_1-x_2\over 2}\right)^2
\end{equation}
which is a direct consequence of the map (\ref{map2}).
This identity implies the excitation sits on a semicircle in the $X,Z$ plane, in the bulk.
The semicircle has radius $(x_1-x_2)/2$ and center at $X=(x_1+x_2)/2$ and $Z=0$.
To locate the excitation on this semicircle, we can specify an angle $\theta$ as follows
\bea
\tan\theta={Z\over X-{x_1+x_2\over 2}}={2\sqrt{p_1^+p_2^+}\over p_1^+-p_2^+}
\label{ForTheta}
\eea
where the last equality is easily obtained from the map (\ref{map2}).
Using double angle trignometric identities it is simple to verify that the angle $\theta$ we have just defined is the angle 
$\theta$ appearing in the map.
\begin{figure}[h]%
\begin{center}
\includegraphics[width=0.5\columnwidth]{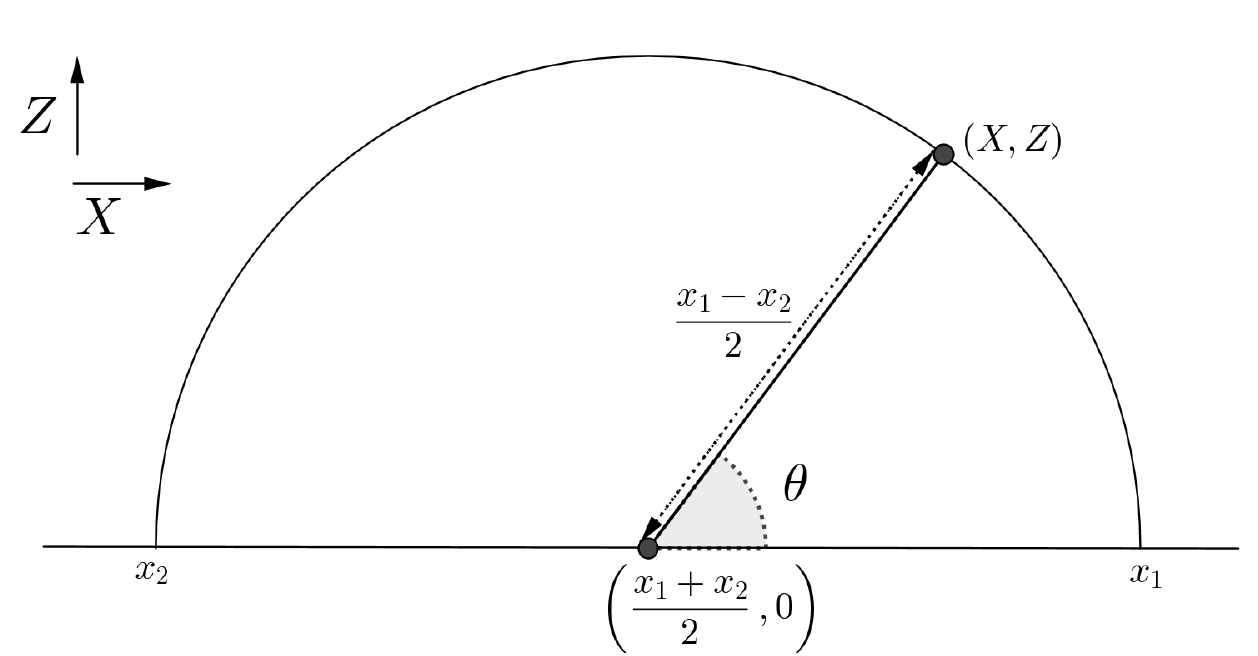}%
\caption{The bilocal describing a pair of excitations localized at $(x_1,p_1^+)$ and $(x_2,p_2^+)$ correspond to a 
bulk excitation localized at $(X,Z)$ as shown.
This figure lives on a constant $x^+=X^+$ slice.
The angle $\theta$ is related to $p_1^+$ and $p_2^+$ according to (\ref{ForTheta}).}
\label{fig:ToBulk}
\end{center}
\end{figure}

If we study excitations confined to a strip subregion of the CFT, defined by $x^+=0$, $-\infty \le x^-\le\infty$ and 
$-L\le x\le L$ then the corresponding bulk excitations can be located at any $\theta,X^-$, but are confined to the
semicircle
\begin{equation}
X^2+Z^2\le L^2\qquad -\infty\le X^-\le\infty\qquad 0\le\theta\le\pi
\end{equation}
of the $(X,Z)$ plane and are at $X^+=0$.
The semicircle bounding the region in which the bulk excitations are located has an attractive interpretation.
The metric of AdS$_4$, written using lightcone coordinates on the Poincare patch, is given by
\bea
ds^2 ={dX^+ dX^- +dX^2+dZ^2\over Z^2}
\eea
The induced metric on a constant $X^+$ slice is thus given by
\bea
ds^2={dX^2+dZ^2\over Z^2}
\eea
Now consider a region ${\cal E}_{\cal R}$ of the $(X,Z)$ plane stretching from the boundary at $Z=0$ to a curve $Z=Z(X)$.
The area of this region is given by
\bea
{\cal A} =\int dX {\sqrt{Z^{\prime\, 2}+1}\over Z}\qquad Z'={dZ\over dX}
\eea
Minimizing the area we find\footnote{Instead of minimizing this area functional, we have used the usual trick of writing
down the conserved quantity implied by the fact that ${\cal A}$ is independent of $X$. We have set the conserved
quantity equal to $R^{-1}$ which is constant. This is why we obtain a first order equation.}
\bea
{dZ\over dX}={\sqrt{R^2-Z^2}\over Z}
\eea
which is solved by $X^2+Z^2=R^2$.
This demonstrates that bulk excitations are restricted to region ${\cal E}_R$ bounded by the boundary of AdS and an 
extremal surface.
Thus, ${\cal E}_{\cal R}$ is nothing but the entanglement wedge!

The above result suggests that we are recovering the entanglement wedge reconstruction.
This conclusion is however, too hasty: recall that the bulk field $\Phi(X^+,X^-,X,Z,\theta)$ defined in the bilocal map
is a mixture of fields of any even spin.
The bulk fields reconstructed from the CFT subregion do not allow us to distinguish these different spin components.
To obtain fields of a definite spin we need to compute the integral
\bea
\int_0^\pi d\theta\,\, \Phi(X^+,X^-,X,Z,\theta)\cos (2s\theta)\label{sIntegral}
\eea
As $\theta\to 0$ we must have ${x_1+x_2\over 2}\to -\infty$ and to get $\theta=\pi$ we must have 
${x_1+x_2\over 2}\to \infty$.
Thus, as $\theta$ ranges from $0$ to $\pi$, $x$ ranges over the entire boundary.

In conclusion, even to reconstruct the complete set of higher spin fields at just a single point in the bulk, we require the 
entire boundary.
This is not unexpected: since AdS/CFT is an isomorphism between two theories, it is highly unlikely that a given subregion 
of the CFT is dual to a given subregion of the bulk spacetime.
Within bilocal holography, we see explicitly that this is not the case: AdS/CFT provides only a global map between bulk 
and boundary states.

\section{Code Subspace and Quantum Error Correction}\label{qecsec}

A crucial ingredient in proposals for the role of quantum error correction in holography and entanglement wedge 
reconstruction, are that we restrict ourselves to the code subspace of the full Hilbert space.
The code subspace provides a correct description of effective field theory around a given classical background. 
Any given observer can not access the complete Hilbert.
For example, they can't access infinite energy states and can only act with a subset of all the fields in the theory.
This is what the code subspace accounts for.
The code subspace is closely related to how the little Hilbert space of states, nearby a reference state, is built in the 
work of Papadodimas and Raju \cite{Papadodimas:2012aq,Papadodimas:2013jku}.

\subsection{Quantum Error Correction}

We will follow the prescription for the code subspace given in \cite{Almheiri:2014lwa}.
One starts by choosing a finite set of local bulk operators $\varphi_i(X^\mu)$, realized in the CFT using the 
representation provided by bilocal holography. 
The code subspace ${\cal H}_C$ is the linear span of states of the form
\bea
|\Omega\rangle,\qquad \varphi_i(X^\mu)|\Omega\rangle,\qquad 
\varphi_i(X^\mu_1)\varphi_j (X^\mu_2)|\Omega\rangle,\cdots
\eea
$|\Omega\rangle$ is the ground state of the system.
The range of $i$, which determines the number of fields we act with and the number of points where the operators are 
located, is bounded by a fixed finite number.

Restricting to the code subspace has far reaching consequences.
By assumption, at any given bulk point, we have a linear combination of a finite number of AdS$_4$ fields, with known spins.
To separate the components of a given spin we don't need to use (\ref{sIntegral}) so that $\theta$ need not 
run continuously from $0$ to $\pi$.
Given a sum of a finite number of fields of known spins, we can solve for each component given the sum at 
a finite number of $\theta$'s.
By choosing different bilocal fields, we obtain distinct semicircles giving the value of the bulk field $\Phi(X^+,X^-,X,Z,\theta)$ 
for a sufficiently large number of values of $\theta$.
An example is shown in Figure \ref{fig:FinTheta} below.
Consequently from a given subregion ${\cal R}$ of the CFT we can reconstruct the fields that belong to the code subspace 
in the bulk region corresponding to the entanglement wedge ${\cal E}_{\cal R}$.
\begin{figure}[h]%
\begin{center}
\includegraphics[width=0.5\columnwidth]{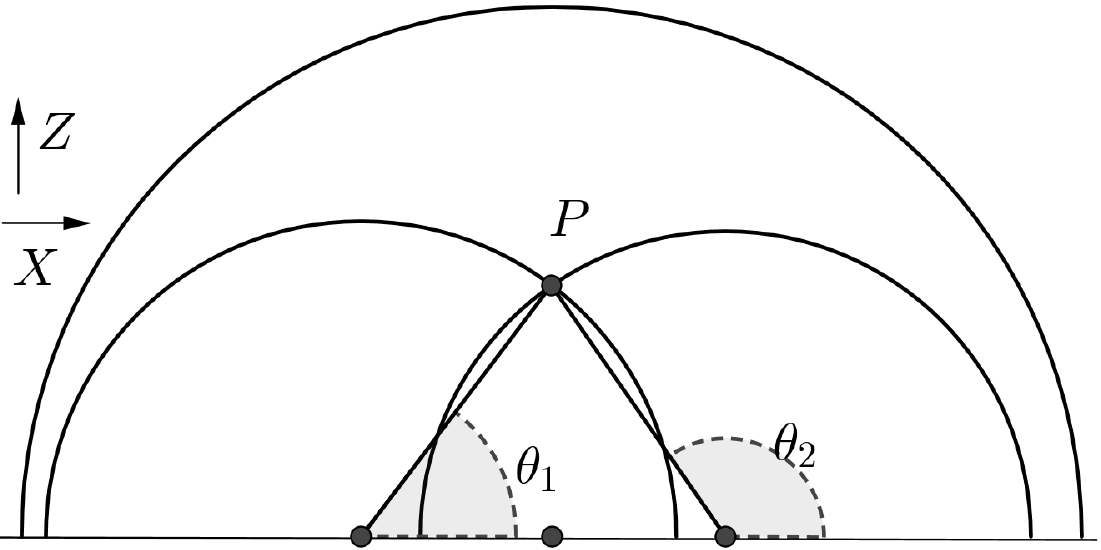}%
\caption{It is possible to choose distinct bilocals from the subregion ${\cal R}$ that correspond to different semicircles passing
through point $P$. This allows us to obtain the value of the bulk field $\Phi(X^+,X^-,X,Z,\theta)$ at $P$ for a number
of distinct $\theta$.}
\label{fig:FinTheta}
\end{center}
\end{figure}

The properties that motivated the quantum error correction formulation of holography are now easy to demonstrate.
To see how a single bulk operator corresponds to multiple boundary operators, constructed using different subregions,
consider Figure \ref{multiple} below.
In subregion $A$ we reconstruct the values of $\Phi(X^+,X^-,X,Z,\theta)$ at angles $\theta_i <{\pi\over 2}$, while in 
subregion $B$ we reconstruct at angles  $\theta_i >{\pi\over 2}$.
This demonstrates an ambiguity in the bulk reconstruction, since the bulk field is coded into different subregions of the CFT.  
\begin{figure}[h]%
\begin{center}
\includegraphics[width=0.5\columnwidth]{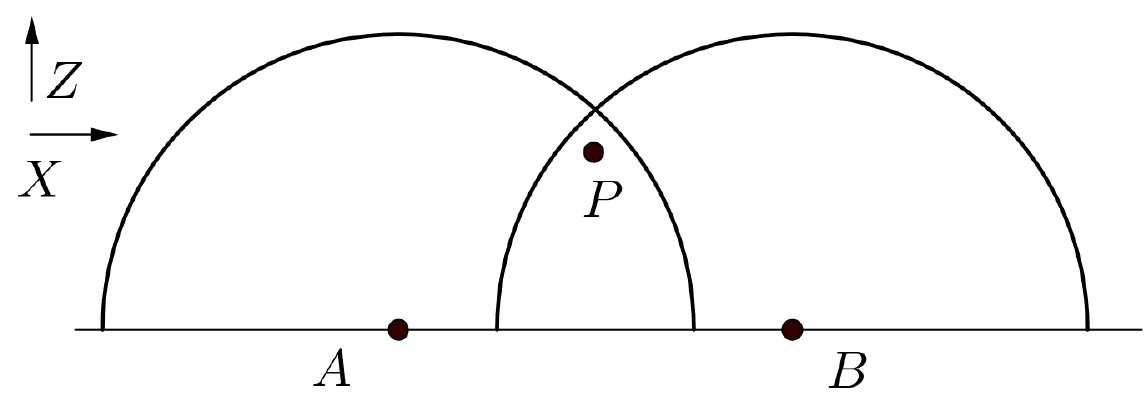}%
\caption{The fields at point $P$ can be reconstructed using bilocals taken from region $A$ or region $B$.
It is however not possible to reconstruct the bulk fields at point $P$ using bilocals from the intersection $A\cap B$.}
\label{multiple}
\end{center}
\end{figure}

\begin{figure}[h]%
\begin{center}
\includegraphics[width=0.5\columnwidth]{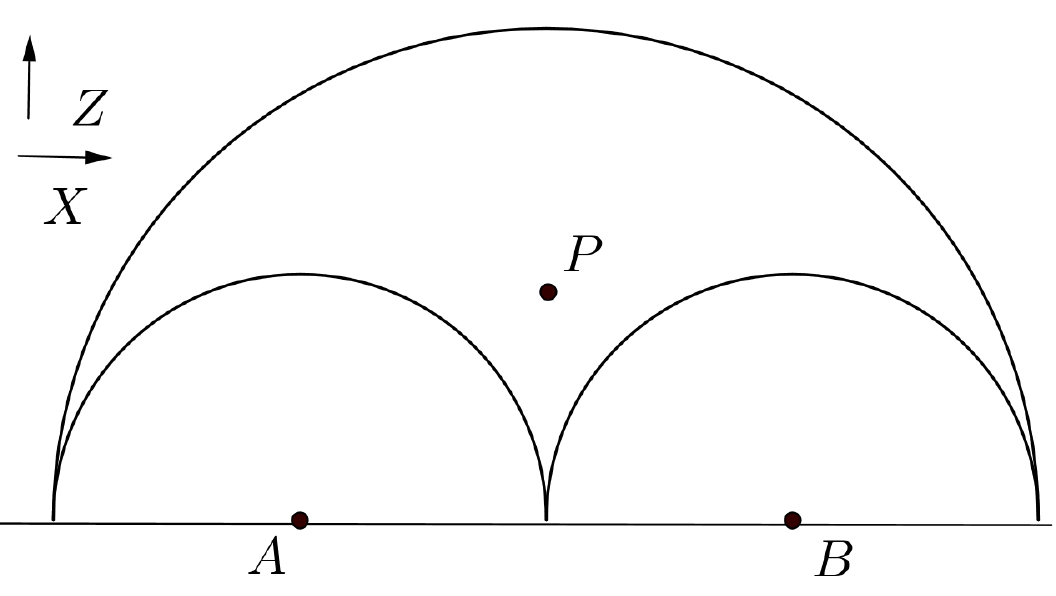}%
\caption{It is not possible to reconstruct the bulk operator at $P$ if we restrict to either subregion $A$ or $B$.
The reconstruction is however possible if bilocals from $A\cup B$ are used.}
\label{fig:QECSolid}
\end{center}
\end{figure}

From Figure \ref{fig:QECSolid} below it is clear the bulk field at the point $P$ can not be reconstructed using either of the
subregions $A$ or $B$. 
However, if bilocals from the union $A\cup B$ are used, we can construct the bulk field at point $P$.
The new bilocals, not present in either $A$ or $B$, are bilocals with one point in $A$ and one in $B$.
These bilocals are separated enough that they reach deep enough into the bulk to reconstruct $P$; they represent
information that is shared non-locally between the two boundary subregions.

\subsection{Entanglement Entropy and Bit Threads}

Another observable, associated to a subregion $A$ of the CFT, is the entanglement entropy $S(A)$.
An important formula describing this observable is the RT formula, which states that entanglement 
entropy of a region in the boundary CFT is determined by a geometric extremization problem in the bulk 
\cite{Ryu:2006bv,Ryu:2006ef}. 
The RT formula states that the entropy of a spatial region $A$ on the boundary CFT is given by
\bea
S(A)={1\over 4G_N}{\rm area}(m(A))
\eea
where $m(A)$ is a minimal hypersurface in the bulk, homologous to $A$. 
The RT formula holds for a conformal field theory dual to a state represented by a classical spacetime with a 
time-reflection symmetry. 
Since our bulk spacetime is time independent AdS$_4$, we are always within this regime of validity.
Denote a Cauchy slice in the geometry by ${\cal M}$.
We can ensure that ${\cal M}$ is a compact manifold with a boundary by introducing a cutoff ``near'' the conformal boundary.
The CFT lives on the boundary of ${\cal M}$.

The Riemannian version of the max flow-min cut theorem can be used to rewrite the RT formula in terms of the maximal
flow of a vector field $v^\mu$, through any surface homologous to $A$\cite{Freedman:2016zud}.
(See \cite{Headrick:2017ucz} for a nice discussion of the proof of the required theorem.)
The integral curves of any vector field are oriented and locally parallel. 
Following \cite{Cui:2018dyq} it is useful to generalize the notion of integral curves of a vector field, 
by dropping these two conditions.
This gives the concept of a bit thread: bit threads are unoriented curves, that pass through a given neighborhood, 
at different angles and are even allowed to intersect. 
A thread configuration is a set of unoriented curves on ${\cal M}$, such that threads end only on the boundary of 
${\cal M}$ and the thread density is nowhere larger than $1/4G_N$.

How should we think about bit threads?
In what follows we argue that bit threads have a natural interpretation within bilocal holography.
The bilocal field creates gauge invariant excitations, which are naturally entangled as a result of the sum over the
color index
\bea
\sigma(x^+,x_1^-,x_1,x_2^-,x_2)|0\rangle = \sum_{a=1}^N\phi^a(x^+,x_1^-,x_1)\phi^a(x^+,x_2^-,x_2)|0\rangle
\eea
A gauge invariant pair of modes in the vacuum is naturally entangled in this way.
The entanglement entropy of subregion $A$ receives contributions from entangled pairs, which entangle a mode in $A$
with a mode in its complement $\bar{A}$.
Consider a pair of modes which are entangled, with one of the pair localized at position $(x_1^-,x_1)$ in $A$ and the other
at $(x_2^-,x_2)$ in $\bar{A}$.
This entangled pair will have contributions from all possible $p_1^+$ and $p_2^+$ values.
Consequently, the bilocal map implies that the bulk degree of freedom is smeared over the semicircle with 
center ${1\over 2}(x_1+x_2)$ and radius ${1\over 2}(x_1-x_2)$.
Thus, in the bulk these entangled pairs are represented by semicircles which end at the two modes being entangled.
The semicircles indicate where the corresponding bulk modes are smeared.
This is precisely what a bit thread is: the bit threads connecting $A$ to $\bar{A}$ are manifestations of the entanglement
between $A$ and $\bar{A}$. 
We can visualize each bit thread as a ``channel'' allowing a bit of (quantum) information to be communicated 
between different regions on the spatial boundary. 
Recall that a thread configuration, by definition, is a set of unoriented curves on ${\cal M}$, such that threads end only on 
the boundary of ${\cal M}$ and the thread density is nowhere larger than $1/4G_N$.
The semicircles of the bilocal holography, which end on points in the CFT clearly do end on the boundary of ${\cal M}$.
In addition, since these threads end on degrees of freedom in the CFT, the density of threads is naturally restricted by the
holographic bound which dictates that the boundary theory has only one bit of information per Planck area.

As a final piece of evidence in favor of our interpretation, the paper \cite{Agon:2018lwq} constructs explicit examples 
of flows $v^\mu$ for a number of geometries including empty AdS. 
The fact that the family of integral curves is given by a two parameter family of semicircles is in harmony with
our interpretation\footnote{Our geometry obtained by setting $x^+=$ constant is naturally related to the case of
AdS$_3$ of \cite{Agon:2018lwq}. Choosing an entangling regions given by an interval is then naturally related to the 
sphere discussion of Section 2.1.1 of \cite{Agon:2018lwq}.}.

This bit thread perspective gives an attractive and intuitive interpretation to the RT formula.
As an example of the utility of the bit thread perspective\cite{Freedman:2016zud,Cui:2018dyq} recall that
given a disjoint spatial region $A$ of the boundary CFT, the minimal hypersurface homologous to $A$ will not always 
vary continuously even as $A$ is varied continuously.
Bit threads do vary continuously as a function of $A$.
It is satisfying to see the accord between bit threads and how a gauge invariant entangled pair is represented in the
bulk using bilocal holography.

\subsection{Entanglement Wedge Reconstruction}

We end this section considering the problem of entanglement wedge reconstruction, within bilocal holography, in
the empty AdS geometry.
The problem is to construct any bulk operator $\Phi_a$ supported in the entanglement wedge $a$ of $A$ from
boundary observables $O_A$, living on the subregion $A$ of the CFT.
Building on earlier work \cite{Jafferis:2015del} showing the equality of bulk and boundary relative entropies a rather 
precise and elegant characterization of entanglement wedge reconstruction was developed in \cite{Cotler:2017erl}.
For an arbitrary state, we restrict to a subregion by tracing over the complement to this subregion.
Given the equality of bulk and boundary relative entropies, a natural way forwards is to phrase entanglement wedge
reconstruction as a universal recovery channel that undoes the effect of the partial trace.
It is non-trivial \cite{Cotler:2017erl} to prove that there is a universal recovery channel that can recover from the noise 
introduced by restricting from the entire bulk to subregion $a$.
In the case of bilocal holography, things are simpler, as a consequence of the fact that the map from CFT
bilocals $\eta(x^+,p_1^+,x_1,p_2^+,x_2)$ to bulk fields $\Phi(X^+,X^-,X,Z,\theta)$ given in (\ref{niceJ}) is so simple.
A short computation shows
\bea
\eta(x^+,p_1^+,x_1,p_2^+,x_2)&=&\int_{-\infty}^\infty dx^- {e^{i(p_1^++p_2^+)x^-}\over 2\sqrt{p_1^+p_2^+}}\cr
&\times& \Phi\left(x^+,x^-,{p_1^+x_1+p_2^+x_2\over p_1^++p_2^+},
{\sqrt{p_1^+p_2^+}(x_1-x_2)\over p_1^++p_2^+},2\tan^{-1}\sqrt{p_2^+\over p_1^+}\right)
\eea
Now, by convention take $x_1>x_2$.
If we set
\bea
\alpha={p_1^+\over p_1^++p_2^+}\qquad
\beta={p_2^+\over p_1^++p_2^+}
\eea
then we have $\alpha+\beta=1$ and $0\le \alpha,\beta\le 1$.
The $Z=\sqrt{\alpha\beta}(x_1-x_2)$ coordinate is zero when either $\alpha=0$ or $\beta=0$, and it takes a maximum
value of ${1\over 2}(x_1-x_2)$ at $\alpha={1\over 2}=\beta$.
The $X=\alpha x_1+\beta x_2$ coordinate ranges from $x_2$ when $\beta=1$ and $\alpha=0$, to $x_1$ when 
$\alpha=0$ and $\beta=1$.
This proves that a bilocal with points $x_1,x_2$ inside $A$ is reconstructed using bulk fields within $a$.

\section{Non-localities from large $N$}\label{HI}

The modes across the horizon of a black hole must be entangled if the horizon geometry is to be smooth.
Further, for old black holes the Page curve implies that modes just outside the hole must be entangled with
modes that were emitted earlier and are now far from the hole. 
Given that entanglement is monogamous, this represent a paradox \cite{Almheiri:2012rt,Almheiri:2013hfa,Marolf:2013dba} 
(see also \cite{Mathur:2009hf}).
Three possible resolutions to this paradox are
\begin{itemize}
\item[1.] Fuzzballs \cite{Mathur:2005zp,Skenderis:2008qn,Mathur:2012jk,Avery:2012tf}.
\item[2.] Firewalls \cite{Almheiri:2012rt,Almheiri:2013hfa,Marolf:2013dba}.
\item[3.] Holography of information \cite{Papadodimas:2012aq}.
\end{itemize}
The first proposal resolves the paradox by cutting spacetime off at the black hole horizon, so that there are no modes 
inside the horizon.
The second proposal relaxes entanglement between modes just inside and outside the hole, at the cost of large energy 
densities at the horizon, i.e. a wall of fire.
The third proposal resolves the paradox by identifying modes inside the hole with the radiation modes 
in the distant region, far from the hole.
In this section we explain how bilocal holography provides a mechanism to realize the holography of information scenario.
 
According to the quantum error correcting framework for holography, bulk information is coded into boundary degrees 
of freedom, in a redundant way.
Could this redundancy explain why degrees of freedom inside the black hole are identified with degrees of freedom
of the radiation far from the hole?
Probably not.
The redundancy in the quantum error correcting description arises both because information about the bulk is encoded non-locally
into the boundary and because we are focusing on a subspace, the code subspace, that a low energy observer has access to.
The holography of information is a property of the full unitary evolution in a black hole background, using the complete Hilbert 
and not the code subspace.
To understand the holography of information, one has to explain non-local redundancies present in the complete bulk 
Hilbert space.
Where do these non-localities come from?

The central hypothesis of bilocal holography is that a gauge fixed version of the gravitational dual is obtained by a change
to gauge invariant variables in the CFT.
We will argue that the CFT does not have enough independent degrees of freedom to produce a local field theory in
higher dimensions, so that the holographic theory necessarily has non-local redundancies.
We will show that these redundancies imply identifications between degrees of freedom at the boundary and degrees of 
freedom that can be arbitrarily deep in the AdS bulk, which is very reminiscent the holography of information. 
Both the CFT and the higher dimensional gravitational theory have an infinite number of degrees of freedom, making
a comparison difficult. 
To make the discussion well defined, discretize space $(x^-,x)\to(x^-_i,x_i)$ to obtain a lattice with $L$ sites.
The original fields $\phi^a(x^+,x^-_i,x_i)$ give at most $NL$ independent\footnote{Gauge invariance will reduce the number
of physical degrees of freedom to less than $NL$ so this upper bound is strict.} degrees of freedom, while the bilocal 
$\sigma(x^+,x^-_i,x_i,y^-_j,y_j)$ gives $L^2$ degrees of freedom. 
For $L>N$ not all the degrees of freedom in the bilocal can be independent.
These redundancies will show up in complicated states, with more than $\sim 2N$ modes excited to produce the state.
Clearly, this is a finite $N$ effect.
An example of a complicated state is the state of an old black hole, which has an atmosphere of excited quanta 
produced by Hawking radiation.

We can describe these redundancies in enough detail to see that they imply identifications between degrees of freedom deep 
in the bulk and degrees of freedom located close to the boundary.
To simplify the discussion, switch to the mixed position/momentum description, evaluate all bilocals at the same momentum
$p_1^+=p_2^+=p^+$ and discretize the spatial coordinate transverse to the light cone $\sigma(x^+,p^+,x_i,p^+,y_j)$.
The simplest case is when $N=1$. 
In this case it is clear that
\begin{equation}
\sigma(x^+,p^+,x_1,p^+,y_1)\sigma(x^+,p^+,x_2,p^+,y_2)=
\sigma(x^+,p^+,x_1,p^+,y_2)\sigma(x^+,p^+,x_2,p^+,y_1) 
\end{equation}
This constraint between the bilocal degrees of freedom can be written as
\begin{equation}
\det M=0\qquad
M=\left[
\begin{matrix}
\sigma(x^+,p^+,x_1,p^+,y_1) &\sigma(x^+,p^+,x_1,p^+,y_2)\\
\sigma(x^+,p^+,x_2,p^+,y_1) &\sigma(x^+,p^+,x_2,p^+,y_2)
\end{matrix}\right]
\end{equation}
which simply expresses the fact that the $2\times 2$ matrix $M$ has rank 1.
In the general case where we have $N$ components, this constraint can be written as
$\det M=0$ where
\begin{equation}
M=\left[
\begin{matrix}
\sigma(x^+,p^+,x_1,p^+,y_1) &\sigma(x^+,p^+,x_1,p^+,y_2) &\cdots &\sigma(x^+,p^+,x_1,p^+,y_{N+1})\\
\sigma(x^+,p^+,x_2,p^+,y_1) &\sigma(x^+,p^+,x_2,p^+,y_2) &\cdots &\sigma(x^+,p^+,x_2,p^+,y_{N+1})\\
\vdots                                          &\vdots                                           &\ddots &\vdots \\
\sigma(x^+,p^+,x_{N+1},p^+,y_1) &\sigma(x^+,p^+,x_{N+1},p^+,y_2) &\cdots
&\sigma(x^+,p^+,x_{N+1},p^+,y_{N+1})
\end{matrix}\right]
\end{equation}
expressing the fact that the $N+1\times N+1$ dimensional matrix $M$ has rank $N$.
\begin{figure}[h]%
\begin{center}
\includegraphics[width=0.9\columnwidth]{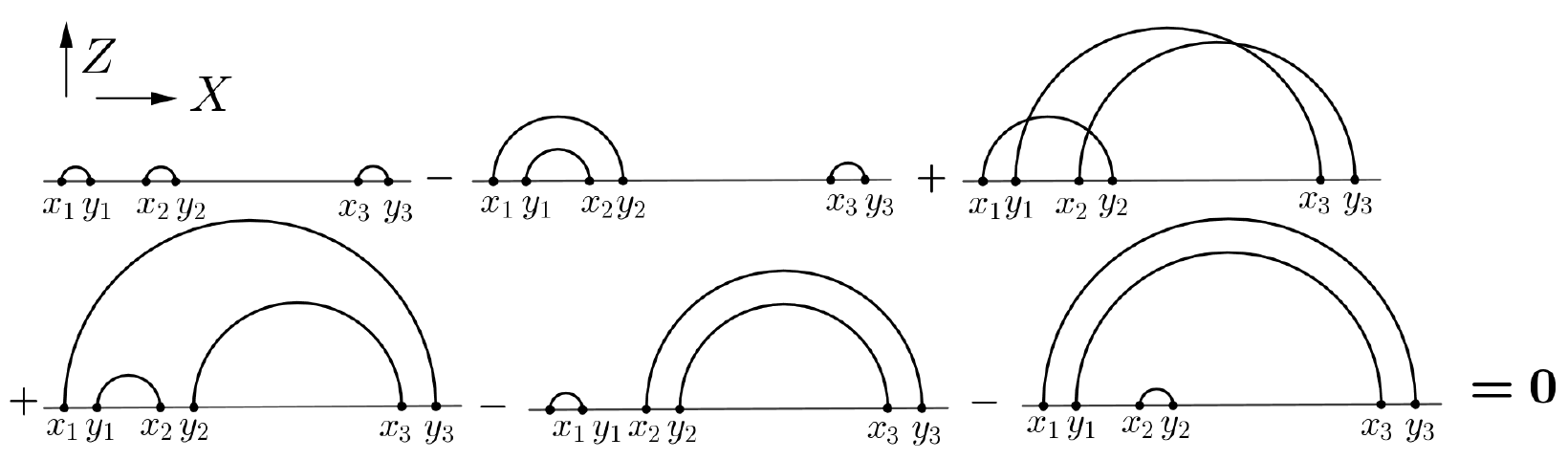}%
\caption{Constraints that arise when $N=2$.}
\label{fig:Nis2}
\end{center}
\end{figure}

We can study the above constraint for some interesting choices of the points $x_i$ and $y_i$.
$N=2$ is already an instructive example.
Choose four points $x_i$ and $y_i$ for $i=1,2$ clustered in a small region and two more points for $i=3$, close to each 
other, but well separated from the first four points. 
The constraint, written diagrammatically, is shown in Figure \ref{fig:Nis2}.
The bulk operator corresponding to each bilocal is located on the semicircle shown.
Since all light cone momenta are equal $\theta=\pi$ and each bulk operator is located on the semicircle at the point which is
deepest in the bulk.
The first two diagrams in Figure \ref{fig:Nis2} correspond to three operators located near the boundary.
Choosing the points $x_1,x_2,y_1,y_2$ to be arbitrarily close to each other, and choosing the points $x_3,y_3$ to be 
arbitrarily close to each other, these operators are arbitrarily close to the boundary.
The remaining four diagrams correspond to two operators located deep in the bulk and an operator located near the boundary.
Making the separation between location of the four points $x_1,x_2,y_1,y_2$ and the location of the two points $x_3,y_3$
arbitrarily large, the operators that explore the bulk are arbitrarily deep in the bulk.
This demonstrates non-local identifications between degrees of freedom at the boundary and degrees of freedom arbitrarily 
deep in the AdS bulk. 
\begin{figure}[h]%
\begin{center}
\includegraphics[width=0.9\columnwidth]{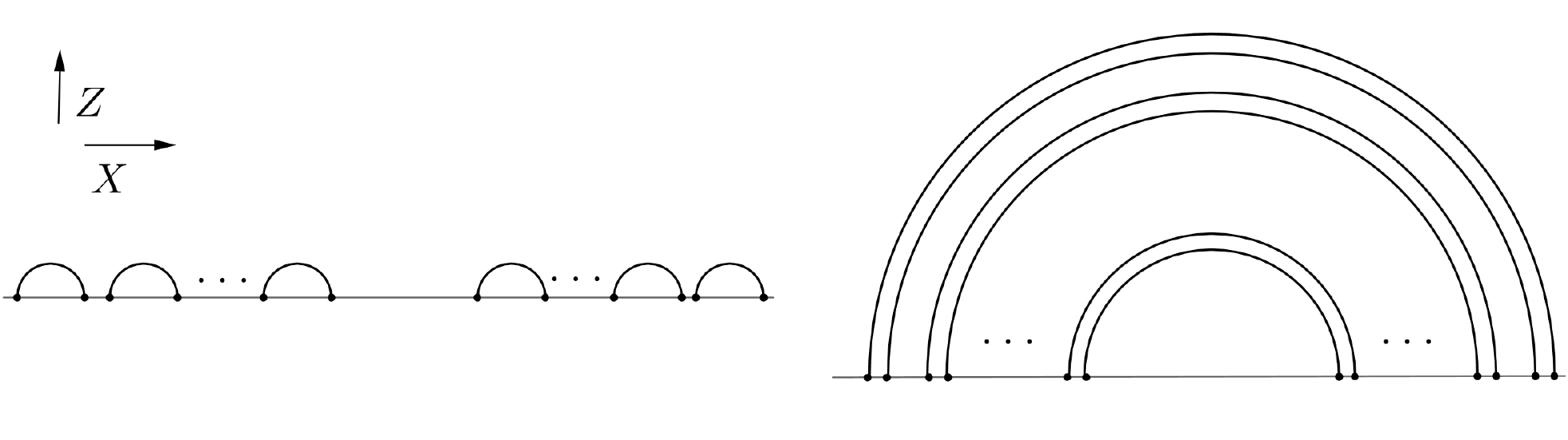}%
\caption{Two of the diagrams participating in the constraint that arises for general $N$.}
\label{fig:N}
\end{center}
\end{figure}

For general $N$ the relations are more complicated, but the conclusion is unchanged.
For example, two of the diagrams participating in the constraint that arises from considering $N+1$ pairs of points are
shown in Figure \ref{fig:N}.
The first diagram is again associated with degrees of freedom that are localized at the boundary whilst the second describes
degrees of freedom localized arbitrarily deep in the bulk.

\section{Conclusions}\label{conclusions}

Bilocal holography gives an explicit mapping between the degrees of freedom of a CFT and its gravitational dual.
We have discussed the example of bilocal holography, relating the free $O(N)$ vector model and it's dual higher spin
gravity in AdS$_4$ spacetime.
The map between degrees of freedom is one-to-one at large $N$ and there is no ambiguity in the map.

To describe effective field theory relevant for a low energy observer, one introduces a code subspace.
The code subspace accounts for the fact that a low energy observer only has access to a subspace of the full Hilbert space.
There is a redundancy in how the states of the code subspace are encoded in the CFT and in this way we can see how the
quantum error correction description of holography is recovered from bilocal holography.
Further, by inverting the operator mapping we prove entanglement wedge reconstruction.
The bilocal of the boundary theory maps into a line in the bulk with endpoints given by the bilocal coordinates, in a
way that is reminiscent of the bit threads description of entanglement.

At finite $N$ bilocal holography predicts a genuine redundancy of the gravitational degrees of freedom: complicated
states in the gravity dual can have more degrees of freedom excited than there are independent degrees of freedom in the
CFT. In this case, there will be relations between naively independent degrees of freedom in the gravitational theory.
We gave a rudimentary analysis of this issue and argued that the relations which appear naturally relate degrees of 
freedom localized at the boundary with degrees of freedom deep in the bulk of AdS.
This has important implications for how information is coded into the gravitational theory.

There are a number of ways in which the analysis presented in this paper can be extended.
A simple extension would be to repeat the analysis using the canonical map developed in \cite{Koch:2014aqa}.
Another interesting extension would be to explore the Wilson-Fisher fixed point conformal field theory.
For results in this direction see \cite{Mulokwe:2018czu}.
A much more interesting analysis would be to apply the ideas of bilocal holography (i.e. of collective field 
theory \cite{Jevicki:1979mb,Jevicki:1980zg}) to theories of matrices.
The number and type of gauge invariants that can be constructed is much richer, but the ideas would be the same: one
first changes to invariant variables and then defines a coordinate transformation that will naturally adapt the generators
to the direct sum of representations that appear in a given multi-local gauge invariant.

\section*{Acknowledgement}
We would like to thank Junggi Yoon for useful discussions. 
This work is supported by the Science and Technology Program of Guangzhou (No. 2019050001 and
No. 2020A1515010388), by the National Natural Science Foundation of China under Grant No. 12022512 and No. 12035007,
by a Simons Foundation Grant Award ID 509116 and by the South African Research Chairs initiative 
of the Department of Science and Technology and the National Research Foundation.

\begin{appendix}

\section{Coordinate Transformation}\label{BMap}

An important component of bilocal holography is the relation between the CFT and AdS coordinates given in 
(\ref{map1}) and (\ref{map2}).
This mapping of the coordinates is determined entirely by the symmetries of the problem as we explain in this section.
The basic idea is the following: each scalar field $\phi^a(x)$ transforms in the short representation $V_{\Delta,s}$ 
of $SO(2,3)$ constructed on the primary of dimension $\Delta={1\over 2}$ and spin $s=0$.
The bilocal transforms in the tensor product $V_{{1\over 2},0}\otimes V_{{1\over 2},0}$, which is reducible
\begin{equation}
V_{{1\over 2},0}\otimes V_{{1\over 2},0} = V_{1,0}\,\oplus\,\,\bigoplus_{s=2,4,6,\cdots} V_{s+1,s}
\label{FlatoFronsdal}
\end{equation}
The original bilocal coordinates are the natural set of coordinates to describe the tensor product on the LHS of
(\ref{FlatoFronsdal}), while the AdS coordinates are natural to describe the RHS.

We will start with a trivial toy model to illustrate the idea and then verify the origin of the transformation (\ref{map1}).

\subsection{Motivational Example}

Consider the tensor product of two spin ${1\over 2}$-particles. 
This decomposes into the triplet and the singlet
\begin{equation}
{1\over 2}\otimes {1\over 2} = 1\oplus 0\label{twospins}
\end{equation}
In the basis natural for the LHS of (\ref{twospins}), we have the $z$ component of spin (for example) given by
the standard co-product
\begin{equation}
J^z= \Delta({1\over 2}\sigma^3)={1\over 2}\sigma^3\otimes {\bf 1}+{\bf 1}\otimes{1\over 2}\sigma^3
\end{equation}
where ${\bf 1}$ is a $2\times 2$ identity matrix and $\sigma^3$ is a Pauli matrix.
In the basis natural for the RHS of (\ref{twospins}), the $z$ component of spin is given by
\begin{equation}
J^z= J^z_1\oplus J^z_0
\end{equation}
where $J^z_1$ is a $3\times 3$ matrix with eigenvalues equal to $-1,0,1$ and $J^z_0$ is a $1\times 1$ matrix with
eigenvalue equal to 0.
In moving from the basis which is natural for the LHS of (\ref{twospins}) to the basis which is natural for the RHS, we see
that the generators are transformed.
We will see exactly the same features in the vector model.
The transformation between the bases natural for the two sides of (\ref{twospins}) is accomplished by the change of
coordinates given in (\ref{map1}) and (\ref{map2}).

\subsection{Conformal Transformations}

The generators of conformal transformations for the bilocal $\sigma (x^+,x_1^-,x_1,x_2^-,x_2)$ are obtained using
the standard coproduct 
\begin{equation}
\Delta (L)=L\otimes 1+1\otimes L
\end{equation}
where $L$ is a generator acting on the field $\phi^a(x^+,x^-,x)$.
This is the direct product representation i.e. the analog of the LHS of (\ref{twospins}) and it is the representation
naturally derived from the CFT.
A complete basis for so$(2,3)$ is
\begin{eqnarray}
P^+&=&p_1^++p_2^+\cr\cr
P^x&=&{\partial\over \partial x_1}+{\partial\over\partial x_2}\cr\cr
P^-&=&-{1\over 2p_1^+}{\partial^2\over\partial x_1^2}-{1\over 2p_2^+}{\partial^2\over\partial x_2^2}\cr\cr
J^{+-}&=&x^+ P^- +{\partial\over\partial p_1^+}\, p_1^++{\partial\over\partial p_2^+}\, p_2^+\cr\cr
J^{+x}&=&x^+ \left({\partial\over\partial x_1}+{\partial\over\partial x_2}\right)-x_1 p_1^+ -x_2 p_2^+\cr\cr
J^{-x}&=&-{\partial\over\partial p_1^+}{\partial\over\partial x_1}-{\partial\over\partial p_2^+}{\partial\over\partial x_2}
+{x_1\over 2p_1^+}{\partial^2\over\partial x_1^2}+{x_2\over 2p_2^+}{\partial^2\over\partial x_2^2}\cr\cr
D&=&x^+P^- -{\partial\over\partial p_1^+}p_1^+ -{\partial\over\partial p_2^+}p_2^+
+x_1 {\partial\over\partial x_1}+x_2 {\partial\over\partial x_2} + 1\cr\cr
K^+&=&-{1\over 2}\left(-2 x^+ \left({\partial\over\partial p_1^+}p_1^++{\partial\over\partial p_2^+}p_2^+\right)
+x_1^2 p_1^++x_2^2 p_2^+\right)+x^+D\cr\cr
K^-&=& {3\over 2}\left({\partial\over\partial p_1^+}+{\partial\over\partial p_2^+}\right)
+p_1^+ {\partial^2\over\partial p_1^{+\,\,2}}+p_2^+ {\partial^2\over\partial p_2^{+\,\,2}}
-x_1{\partial\over\partial x_1}{\partial\over\partial p_1^+}-x_2{\partial\over\partial x_2}{\partial\over\partial p_2^+} 
+{x_1^2\over 4p_1^+}{\partial^2\over\partial x_1^2}+{x_2^2\over 4p_2^+}{\partial^2\over\partial x_2^2}\cr\cr
K^x&=&-{1\over 2}\left(-2 x^+ \left({\partial\over\partial x_1}{\partial\over\partial p_1^+}
+{\partial\over\partial x_2}{\partial\over\partial p_2^+}\right)+x_1^2 {\partial\over\partial x_1}
+x_2^2 {\partial\over\partial x_2}\right)\cr
&&+x_1 \left(-x^+ {1\over 2p_1^+}{\partial^2\over\partial x_1^2}
-{\partial\over\partial p_1^+}p_1^+ + x_1{\partial\over\partial x_1}+{1\over 2}\right)
+x_2 \left(-x^+ {1\over 2p_2^+}{\partial^2\over\partial x_2^2}
-{\partial\over\partial p_2^+}p_2^+ + x_2{\partial\over\partial x_2}+{1\over 2}\right)\cr\cr
&&
\end{eqnarray}
The analog of the RHS of (\ref{twospins}) is provided by the generators for completely symmetric tensor fields in light cone
gauge in AdS, which have been worked out by Metsaev in \cite{Metsaev:1999ui}.
Notice that this representation, since it is determined by the transformation of the bulk fields, is the natural representation
for the bulk gravitational description.
We now want to argue that the change of coordinates (\ref{map1}) followed by a simple similarity transform, 
takes us from the above generators (natural for CFT) to those of \cite{Metsaev:1999ui} (natural for AdS).
After changing coordinates using (\ref{map1}), we obtain the following generators
\begin{eqnarray}
P^{+} & = & p^{+} \cr\cr
P^{x} & = & \frac{\partial}{\partial x} \cr\cr
P^{-} & = & -\frac{1}{2 p^{+}} \left( \frac{\partial^2}{\partial x^2} + \frac{\partial^2}{\partial z^2}
 \right) \cr\cr
J^{+-} & = & x^{+} P^{-} - \left(-\frac{1}{p^{+}} - \frac{\partial}{\partial p^{+}} \right)p^{+} \cr\cr
J^{+x} & = & - p^{+} x + x^{+} \frac{\partial}{\partial x} \cr\cr
J^{-x} & = & \left(- \frac{\partial}{\partial p^{+}}  - \frac{1}{p^{+}} \right)\frac{\partial}{\partial x} - x P^{-}  + \frac{1}{p^{+}}\left( \frac{\partial}{\partial \theta} + \cot(\theta) \right) \frac{\partial}{\partial z} \cr\cr
D & = & x^{+} P^{-} + x \frac{\partial}{\partial x} + z \frac{\partial}{\partial z} + p^{+}\left(-\frac{\partial}{\partial p^{+}} - \frac{1}{p^{+}} \right) \cr\cr
K^{+} & = &  -\left( x^{+}\left(  -\frac{1}{p^{+}} - \frac{\partial}{\partial p^{+}}\right) + \frac{1}{2} (x^2 + z^2)\right) p^{+} + x^{+} D \cr\cr
K^{-} & = & - \left( x^{+}\left(-\frac{1}{p^{+}}-\frac{\partial}{\partial p^{+}} \right)   + \frac{1}{2}(x^2 + z^2) \right)P^{-} + \left(-\frac{1}{p^{+}} - \frac{\partial}{\partial p^{+}} \right)D  \cr\cr 
& & + \frac{1}{p^{+}}  \left( x\frac{\partial}{\partial z} - z\frac{\partial}{\partial x}\right)\left( \frac{\partial}{\partial \theta} + \cot\theta \right) + \frac{1}{p^{+}}\left( \frac{\partial}{\partial \theta} + \cot\theta \right)^2    \cr\cr
K^{x} &=& - x^{+}\left( -\frac{1}{p^{+}} - \frac{\partial}{\partial p^{+}} \right) \frac{\partial}{\partial x} - \frac{1}{2}(x^2 + z^2) \frac{\partial }{\partial x}  + x D +  \left( \cot\theta + \frac{\partial}{ \partial \theta}\right) z \cr\cr
& &  - \frac{1}{p^{+}}\left( \cot(\theta) + \frac{\partial}{\partial \theta}\right) \frac{\partial}{\partial z} x^{+}
\end{eqnarray}
Now, rescaling each of the generators $L$ as follows $L\to \mu L{1\over\mu}$ where $\mu =2\pi p^+\sin\theta$, the algebra 
is obviously unchanged. 
To understand why this similarity transform is required, note that our map can be written as
\begin{eqnarray}
\Phi(X^+,p^+,X,Z,\theta) &=& 2\pi
p^+\sin\theta\,\, \eta(X^+, p^+\cos^2{\theta\over 2},X+Z \tan{\theta\over 2}, 
p^+\sin^2{\theta\over 2},X-Z\cot {\theta \over 2})
\cr\cr
&=&\mu\,\, \eta(X^+, p^+\cos^2{\theta\over 2},X+Z\tan{\theta\over 2}, 
p^+\sin^2{\theta\over 2},X-Z\cot{\theta\over 2})
\end{eqnarray}
The rescaling is required since
\begin{eqnarray}
\tilde\eta=L\eta\qquad\Rightarrow\qquad\mu\tilde\eta =\mu L {1\over \mu}\mu\eta\qquad
\Rightarrow\qquad\tilde\Phi =\mu L {1\over \mu}\,\,\Phi
\end{eqnarray}
so that if the generator acting on the bilocal $\eta$ is $L$, then the generator acting on $\Phi=\mu\eta$ is $\mu L {1\over \mu}$.
After the rescaling the generators become 
\begin{eqnarray}
P^{+} & = & p^{+} \cr\cr
P^{x} & = & \frac{\partial}{\partial x} \cr\cr
P^{-} & = & -\frac{1}{2 p^{+}} \left( \frac{\partial^2}{\partial x^2} + \frac{\partial^2}{\partial z^2}
 \right) \cr\cr
J^{+-} & = & x^{+} P^{-} - \left(- \frac{\partial}{\partial p^{+}} \right)p^{+} \cr\cr
J^{+x} & = & - p^{+} x + x^{+} \frac{\partial}{\partial x} \cr\cr
J^{-x} & = & \left(- \frac{\partial}{\partial p^{+}}\right)\frac{\partial}{\partial x} - x P^{-}  + \frac{1}{p^{+}}\left( \frac{\partial}{\partial \theta}\right) \frac{\partial}{\partial z} \cr\cr
D & = & x^{+} P^{-} + x \frac{\partial}{\partial x} + z \frac{\partial}{\partial z} + p^{+}\left(-\frac{\partial}{\partial p^{+}}
\right) \cr\cr
K^{+} & = &  -\left( x^{+}\left(-\frac{\partial}{\partial p^{+}}\right) + \frac{1}{2} (x^2 + z^2)\right) p^{+} + x^{+} D \cr\cr
K^{-} & = & - \left( x^{+}\left(-\frac{\partial}{\partial p^{+}} \right)   + \frac{1}{2}(x^2 + z^2) \right)P^{-} + \left(- \frac{\partial}{\partial p^{+}} \right)D  \cr\cr 
& & + \frac{1}{p^{+}}  \left( x\frac{\partial}{\partial z} - z\frac{\partial}{\partial x}\right)\left( \frac{\partial}{\partial \theta} \right) + \frac{1}{p^{+}}\left( \frac{\partial}{\partial \theta}\right)^2    \cr\cr
K^{x} &=& - x^{+}\left(- \frac{\partial}{\partial p^{+}} \right) \frac{\partial}{\partial x} - \frac{1}{2}(x^2 + z^2) \frac{\partial }{\partial x}  + x D +  \left(\frac{\partial}{ \partial \theta}\right) z \cr\cr
& &  - \frac{1}{p^{+}}\left(\frac{\partial}{\partial \theta}\right) \frac{\partial}{\partial z} x^{+}
\label{fingenerators}
\end{eqnarray}
A comparison with the expressions in section 3.8 of \cite{Metsaev:1999ui} shows complete agreement, for all generators
except for $D$, after identifying
\begin{eqnarray}
\begin{matrix}
&\partial^{+}  \longleftrightarrow  p^{+}& &x^{-}  \longleftrightarrow - \frac{\partial}{\partial p^{+}} \\
&   &\\
&M^{xz} \longleftrightarrow \frac{\partial}{\partial \theta}& 
&\qquad\,\,\, M^{i-}\longleftrightarrow - \frac{1}{p^{+}}\frac{\partial}{\partial \theta}\frac{\partial}{\partial z} 
\end{matrix}
\end{eqnarray}
along with $A = 0$. 
The first two relations are simple consequences of the Fourier transform.
To understand the last two relations we need to interpret, for example, the operator $M^{xz}$.
Recall that the special conformal generators generate a local scaling and Lorentz rotation.
Consequently, it must be that $M^{xz}$ is generating a rotation in the $X,Z$ plane. 
It is clear from Figure \ref{fig:ToBulk} that the angle describing rotations in this plane is $\theta$ so that the above
expression for $M^{xz}$ is indeed correct.
The difference between the expression for $D$ in (\ref{fingenerators}) and that in \cite{Metsaev:1999ui} is easily
explained as follows: $D$ in \cite{Metsaev:1999ui} acts on the field $\Phi(X^+,X^-,X,Z,\theta)$ which has dimension
$\Delta =1$. $D$ in (\ref{fingenerators}) acts on the field
\bea
\Phi(X^+,p^+,X,Z,\theta)=\int dX^- e^{ip^+X^-}\Phi(X^+,X^-,X,Z,\theta)
\eea
which has dimension $\Delta=0$.
The requirement that our change of coordinates implements the passage from the direct product representation (natural for 
the CFT) to the direct sum representation (natural for bulk AdS) determines the map (\ref{map1}) and (\ref{map2}).

\section{Other approaches}

A work that is closely related to bilocal holography was given in \cite{Aharony:2020omh}.
The map proposed in \cite{Aharony:2020omh} intertwines a given bulk SO(2,3) representation labeled by
dimension $\Delta$ and spin $J$ in the bulk  with the SO(2,3) representation of the bilocal.
Since intertwining maps are unique, this shows how the representation of SO(2,3) in the bulk and the bilocal
representation in the boundary determine the map of \cite{Aharony:2020omh}.
In this sense it is similar to the bilocal holography map of \cite{Koch:2010cy} which maps the representation
of the bilocal to all relevant bulk representations.

The map of \cite{Aharony:2020omh} is written using the bulk to boundary propagator $G_{\Delta,J}\left(X,P;W,D_Q\right)$
written in embedding space both for the CFT and the AdS, as well as conformal three point functions
$\langle O_{\frac{1}{2}}(P_1)O_{\frac{1}{2}}(P_2)O_{\tilde\Delta,J}(P,Q)\rangle$. 
We refer the reader to the original paper for further explanation of the notation.
To understand the map of \cite{Aharony:2020omh} we use two known facts:
\begin{itemize}
\item[1.] Section 5.1 of \cite{Aizawa:2014yqa} proves
\bea
[L^{\rm bulk}_{\Delta,J}\,\Phi_{J}]\left(X,W\right)&=& \int dP\,\, G_{\Delta,J}\left(X,P;W,D_Q\right) 
[L^{\rm boundary}_{\Delta,J}\,O_{\Delta,J}](P,Q)
\eea
This says that the bulk-to-boundary map intertwines the bulk and boundary representations.
\item[2.] Formula (3.2) of \cite{Dobrev:1976vr} proves
\bea
&&L^{\rm boundary}_{\Delta,J}\, \left(\int dP_{1}dP_{2}
\langle O_{\frac{1}{2}}(P_1)O_{\frac{1}{2}}(P_2)O_{\tilde\Delta,J}(P,Q)\rangle \eta(P_1,P_2)\right)\cr
&&\int dP_{1}dP_{2}
\langle O_{\frac{1}{2}}(P_1)O_{\frac{1}{2}}(P_2)
[L^{\rm boundary}_{\Delta,J}\,O_{\tilde\Delta,J}](P,Q)\rangle \eta(P_1,P_2)\cr
&=&\int dP_{1}dP_{2}
\langle O_{\frac{1}{2}}(P_1)O_{\frac{1}{2}}(P_2)O_{\tilde\Delta,J}(P,Q)\rangle 
[L^{\rm bilocal}\,\tilde\eta](P_1,P_2)
\eea
This says the three point function intertwines the representation $\Delta,J$ and the bilocal representation.
\end{itemize}

It is useful to write the map of \cite{Aharony:2020omh} as ($\tilde{\eta}$ is the shadow of $\eta$)
\bea
\Phi_{J}\left(X,W\right)&=&
\frac{16\pi^d \Gamma^2(\frac{d}{2}-\frac{1}{2})}{\Gamma^2(\frac{1}{2})2N_{\Delta,J}\lambda_{\Delta,J}} \,
\left( \frac{\Gamma(\frac{d-2}{2})}{4\pi^{\frac{d}{2}}}\right)^2
\int_{P.S.}\frac{d\Delta}{2\pi i}\int 
	\frac{dP}{J!\left(\frac{d}{2}-1\right)_{J}}
\int dP_{1}dP_{2}\cr  
&\times& G_{\Delta,J}\left(X,P;W,D_Q\right)
	 \langle O_{\frac{1}{2}}(P_1)O_{\frac{1}{2}}(P_2)O_{\tilde\Delta,J}(P,Q)\rangle
       \tilde\eta(P_1,P_2)
\eea
The map of \cite{Aharony:2020omh} is an intertwining operator, intertwining a $\Delta,J$ representation
of the bulk SO(2,3) with the bilocal representation.
This follows because the map of \cite{Aharony:2020omh} is a composition of the map defined by the three point function
with the map defined by the bulk-to-boundary map.
The proof is immediate using the above two results
\bea
[L^{\rm bulk}_{\Delta,J}\,\Phi_{J}]\left(X,W\right)&=&
\frac{16\pi^d \Gamma^2(\frac{d}{2}-\frac{1}{2})}{\Gamma^2(\frac{1}{2})2N_{\Delta,J}\lambda_{\Delta,J}} \,
\left( \frac{\Gamma(\frac{d-2}{2})}{4\pi^{\frac{d}{2}}}\right)^2
\int_{P.S.}\frac{d\Delta}{2\pi i}\int 
	\frac{dP}{J!\left(\frac{d}{2}-1\right)_{J}}
\int dP_{1}dP_{2}\cr  \cr
&\times& G_{\Delta,J}\left(X,P;W,D_Q\right)
	 [L^{\rm boundary}_{\Delta,J}\langle O_{\frac{1}{2}}(P_1)O_{\frac{1}{2}}(P_2)O_{\tilde\Delta,J}(P,Q)\rangle]
       \tilde\eta(P_1,P_2)\cr\cr\cr
&=&
\frac{16\pi^d \Gamma^2(\frac{d}{2}-\frac{1}{2})}{\Gamma^2(\frac{1}{2})2N_{\Delta,J}\lambda_{\Delta,J}} \,
\left( \frac{\Gamma(\frac{d-2}{2})}{4\pi^{\frac{d}{2}}}\right)^2 \int_{P.S.}\frac{d\Delta}{2\pi i}\int 
\frac{dP}{J!\left(\frac{d}{2}-1\right)_{J}} \int dP_{1}dP_{2}\cr\cr  
&\times& G_{\Delta,J}\left(X,P;W,D_Q\right)
 \langle O_{\frac{1}{2}}(P_1)O_{\frac{1}{2}}(P_2)O_{\tilde\Delta,J}(P,Q)\rangle [L^{\rm bilocal}\,\tilde\eta] (P_1,P_2)\cr
&&
\eea

\end{appendix}

\end{document}